\documentclass[aps,pra,twocolumn,amsmath,amssymb,superscriptaddress]{revtex4-2}
\usepackage{graphicx}
\usepackage{hyperref}
\usepackage{xcolor}

\newcommand{\wzy}[1]{\textcolor{black}{#1}}
\colorlet{BLACK}{black}

\makeatother

\begin{document}
\title{Enhancing Electron-Nuclear Resonances by Dynamical Control Switching}
\author{Sichen Xu}
\affiliation{Key Laboratory of Atomic and Subatomic Structure and Quantum Control (Ministry of Education), and School of Physics, South China Normal University, Guangzhou 510006, China}
\author{Chanying Xie}
\affiliation{Key Laboratory of Atomic and Subatomic Structure and Quantum Control (Ministry of Education), and School of Physics, South China Normal University, Guangzhou 510006, China}
\author{Zhen-Yu Wang}
\email{zhenyu.wang@m.scnu.edu.cn}
\affiliation{Key Laboratory of Atomic and Subatomic Structure and Quantum Control (Ministry of Education), and School of Physics, South China Normal University, Guangzhou 510006, China}
\affiliation{Guangdong Provincial Key Laboratory of Quantum Engineering and Quantum Materials, and Guangdong-Hong Kong Joint Laboratory of Quantum Matter, South China Normal University, Guangzhou 510006, China}
\begin{abstract}
We present a general method to realize resonant coupling between spins
even though their energies are of different scales. Applying the method
to the electron and nuclear spin systems such as a nitrogen-vacancy
(NV) center with its nearby nuclei, we show that a specific dynamical
switching of the electron spin Rabi frequency achieves efficient electron-nuclear
coupling\wzy{, providing a} much stronger quantum sensing signal \wzy{and dynamic nuclear polarization} than previous
methods. This protocol has applications in high-field nanoscale nuclear
magnetic resonances as well as low-power quantum control of nuclear
spins. 
\end{abstract}
\maketitle

\section{Introduction}

\begin{figure}[b]
\centering \includegraphics[width=0.9\columnwidth]{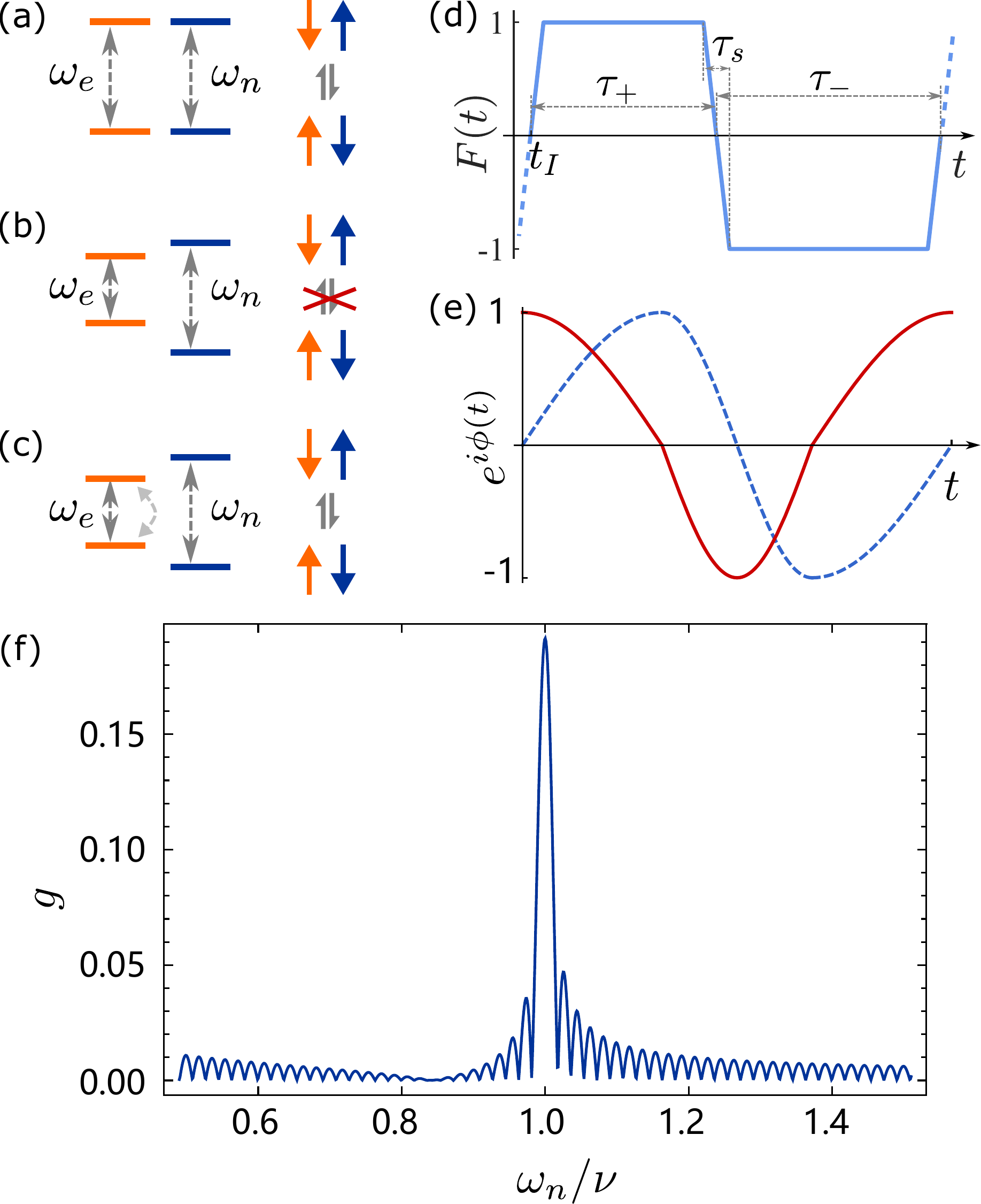} \caption{(a) The spin polarization can exchange when the energies of two spins
(orange and blue arrows) are the same $\omega_{e}=\omega_{n}$. (b)
The exchange of spin polarization is prohibited when $\omega_{e}\protect\neq\omega_{n}$.
(c) Dynamical switching of \wzy{one spin's energy} as $\omega_{e}=F(t)\Omega$
with $F(t)\in\{\pm1\}$ can induce resonant exchange of the spin polarization,
even through $|\omega_{e}|\ll\omega_{n}$. (d) $F(t)$ in one period
$\tau=\tau_{-}+\tau_{+}$ in which the positive control starts at $t=t_I$. (e) The corresponding real (red solid line)
and imaginary (blue dashed line) parts of $e^{i\phi(t)}$ under the
resonance condition Eq.~(\ref{eq:ResonanceEq}) with $k_{D}=1$. The
time integral of $e^{i\phi(t)}$ in one period is nonzero. (f) The
coupling factor $g$ as a function of $\omega_{n}/\nu$ for a total
time $T=50\tau$. Here $\Omega/\nu=0.3$ in all the plots and $\tau_{s}=0$ and $t_{I}=-\frac{1}{2}\tau_{+}$ in (e) and (f).}
\label{fig:Fig01}
\end{figure}

Resonant coupling of spins is a common requirement in spin-based quantum
technologies. A notable platform is the nitrogen-vacancy (NV) center
in diamond~\citep{doherty2013the,dobrovitski2013quantum,schirhagl2014nitrogen},
where the NV electron spin is used to detect, polarize, and control
spins in its vicinity~\citep{bradley2019ten,degen2017quantum,casanova2015robust,glenn2018high,cai2013diamond,wang2016positioning,jiang2022selective}.
Flip-flop dynamics occur when the coupled spins have the same energy,
e.g., due to the same gyromagnetic ratio of the spins {[}see Fig.
\ref{fig:Fig01}(a){]}.

Coupling of nuclear and electron spins is of particular interests.
Electron and nuclear spins usually have quite different energies,
which prohibits their exchange of spin polarization. A dynamical decoupling
(DD) $\pi$ pulse sequence \citep{viola1999dynamical,wang2011protection,souza2012robust,kolkowitz2012sensing,taminiau2012detection,zhao2012sensing,casanova2015robust,munuera2021dynamical,whaites2022adiabatic}
can induce electron-nuclear resonance for quantum information processing
\citep{casanova2016noise,bradley2019ten,tratzmiller2021parallel,pezzagna2021quantum},
nano-scale nuclear magnetic resonance (NMR) \citep{munuera2021dynamical},
quantum sensing \citep{degen2017quantum,nishimura2022floquet}, and
dynamic nuclear polarization (DNP)~\wzy{\citep{jacques2009dynamic,schwartz2018robust,lang2019quantum,tan2019time,corzilius2020high}}.
These DD sequences in the ideal case require unbounded bang-bang control
but in realistic situations the amplitude of control is bounded, where
the latter can leads to spurious resonances between electron and nuclear
spins \citep{loretz2015spurious,haase2016pulse,lang2017enhanced,shu2017unambiguous,wang2019random,wang2020enhancing}.
An alternative scheme is a continuous driving on the electron spin
using a bounded control amplitude, which results in a dressed electron
spin where the energy splitting in the dressed basis equals the Rabi
frequency of the driving \citep{london2013detecting,cai2012robust,cai2013diamond}.
When this Rabi frequency matches the Larmor frequency of the nuclear
spin, i.e., under the Hartmann-Hahn condition \citep{hartmann1962nuclear}
similar to the case in Fig.~\ref{fig:Fig01}(a), the electron and
nuclear spins have resonant coupling.

However, due to the own property of the systems we choose or some
technically limits, sometimes the frequencies between the electron
and nuclear spins have a large mismatch such that the Hartmann-Hahn
condition can not be reached by the Rabi frequency {[}see Fig.~\ref{fig:Fig01}(b){]}.
For instance, under a high magnetic field which would be favorable
to enhance nuclear spin polarization, to prolong spin coherence times,
or to induce large chemical shifts for nano-scale nuclear magnetic
resonance (NMR), the nuclear spin Larmor frequencies can be much larger
than the available Rabi frequency of the electron spin. In some applications
such as those in biological environments \citep{wu2016diamond}, the
maximal control field is restricted to avoid strong microwave heating
effects that could destroy the samples \citep{cao2020protecting}.

In this work, we show that uneven dynamical modulation \wzy{of} the energies can
lead to resonant coupling even \wzy{when} the values of the energies
of the spins have large mismatch values. Our dynamical control switching (DCS) scheme provides more
efficient electron-nuclear spin coupling than recent low-power
control protocols \citep{casanova2019modulated,aharon2019quantum}.
We demonstrate the superior performance of our protocol in quantum
control and sensing of single nuclear spins.

\section{Dynamical switching of energies}

Our theory can be generalized to selectively couple quantum systems of different energy scales. To demonstrate the principle of our protocol, for simplicity, consider two interacting
spins with the Hamiltonian ($\hbar=1$) 
\begin{equation}
H_{0}=H_{S}+H_{I},
\end{equation}
where $H_{S}=\omega_{e}S_{z}+\omega_{n}I_{z}$ is the Hamiltonian
for the two spins with $I_{z}$ and $S_{z}$ being their spin operators.
The interaction $H_{I}=aS_{-}I^{+}+{\rm h.c.}$, where $S_{\pm}=S_{x}\pm iS_{y}$
and $I_{\pm}=I_{x}\pm iI_{y}$, describes the flip-flop dynamics of
the spins. $a$ is the coupling constant. In the interaction picture
of $H_{S}$, the Hamiltonian becomes 
\begin{equation}
\tilde{H}_{I}(t)=aS_{-}I^{+}e^{i(\omega_{n}-\omega_{e})t}+{\rm h.c.},
\end{equation}
if we assume that both the energies $\omega_{e}$ and $\omega_{n}$
are time independent. When $\omega_{e}=\omega_{n}$, the two spins
have resonant flip-flop dynamics with $\tilde{H}_{I}=H_{I}$, see
Fig.~\ref{fig:Fig01}(a). On the other hand, when the energy mismatch
$|\omega_{n}-\omega_{e}|\gg|a|$, the effect of $\tilde{H}_{I}(t)$
is negligible by the rotating-wave approximation, which corresponds
to the case of Fig.~\ref{fig:Fig01}(b).

We aim to modulate the energy mismatch in time to preserve the effect
of the interaction $\tilde{H}_{I}$ even though the energy difference
is large. For simplicity here we only introduce a time dependence
on $\omega_{e}=\omega_{e}(t)$. To see the effect of the modulation
of $\omega_{e}$, we calculate the leading-order effective Hamiltonian
of $\tilde{H}_{I}(t)$ by using the Magnus expansion \citep{mananga2016floquet,haeberlen1968coherent}
for a time $T$; it reads 
\begin{align}
\bar{H} & =\frac{1}{T}\int_{0}^{T}\tilde{H}_{I}(t)dt,\\
 & =gaS_{-}I^{+}+{\rm h.c.}.
\end{align}
The coupling factor 
\begin{equation}
g\equiv\frac{1}{T}\int_{0}^{T}e^{i\phi(t)}dt,\label{eq:g}
\end{equation}
and the difference of the dynamic phase 
\begin{equation}
\phi(t)=\int_{0}^{t}[\omega_{n}-\omega_{e}(t^{\prime})]dt^{\prime}\label{eq:phi}
\end{equation}
can be controlled by $\omega_{e}(t)$. When $\omega_{e}$ is a constant,
the amplitude $g=e^{-ix}\frac{\sin x}{x}$ decays with $x=(\omega_{e}-\omega_{n})T/2$
in a power-law decay manner \citep{haase2018soft}. For the resonant
case $\omega_{e}=\omega_{n}$, $g=1$.

In this work we consider the situation that $\omega_{e}$ is bounded
with its maximal value $\Omega$, with $|\omega_{e}|\leq\Omega\ll\omega_{n}$
and $\omega_{n}-\omega_{e}\gg|a|$. Because $\omega_{n}-\omega_{e}$
is large, the integrand in Eq.~(\ref{eq:phi}) is a fast oscillating
factor, which tends to average out the interaction. To solve the obstacle,
we modulate the dynamic phase $\phi(t)$ with uneven speed to enhance the value of $g$
in Eq.~(\ref{eq:g}). The intuition comes from the idea of uneven modulation of dynamic
phases in quantum adiabatic control which has been used to accelerate quantum adiabatic process
\citep{wang2016necessary,xu2019breaking,zheng2022accelerated,liu2022shortcuts}.
We assume a periodic function $\omega_{e}(t)=\omega_{e}(t+\tau)$,
see Fig.~\ref{fig:Fig01}(d) for an example.  Let the increase of
$\phi(t)$ in one period $\tau$ of $\omega_{e}(t)$ be different
from $2k_{D}\pi$ (with $k_{D}$ an integer) by an amount 
\begin{equation}
\delta_{\phi}=\phi(\tau)-2k_{D}\pi.
\end{equation}
Using the periodicity of $\omega_{e}(t)$ we calculate Eq.~(\ref{eq:g})
for $T=N\tau$, 
\begin{equation}
g=\frac{1}{N\tau}\int_{0}^{N\tau}e^{i\phi(t)}dt=\eta J,\label{eq:g2}
\end{equation}
where the functional $J[\phi(t)]$ is defined as
\begin{equation}
J[\phi(t)]=\frac{1}{\tau}\int_{0}^{\tau}e^{i\phi(t)}dt,
\end{equation}
and the factor 
\begin{equation}
\eta=\frac{1}{N}\sum_{m=1}^{N}e^{i(m-1)\delta_{\phi}}=e^{i\frac{N-1}{2}\delta_{\phi}}\frac{\sin(N\delta_{\phi}/2)}{N\sin(\delta_{\phi}/2)}\label{eq:eta}
\end{equation}
has a peak centered at $\delta_{\phi}=0$ with a width $\sim2\pi/N$,
see Fig.~\ref{fig:Fig01}(f). That is, when $\delta_{\phi}=0$ the
dynamic phase factors in Eq.~(\ref{eq:eta}) coherently add up, while
for other $\delta_{\phi}<\pi$, $\eta\approx0$ for a large averaging
time $T$. 

We want to maximize the functional $J[\phi(t)]$ by using inhomogeneous changes of $\phi(t)$. To have a higher degree of inhomogeneity, we choose 
\begin{equation}
\omega_{e}(t)=F(t)\Omega
\end{equation}
with the periodic modulation function $F(t)=F(t+\tau)$. In particular, we periodically
switch the value of $\omega_{e}$ between the maximal value  $+\Omega$
and the minimal value $-\Omega$. 
We would like to have $F(t)=1$ when $t\in[t_{I},t_{I}+\tau_{+})$ 
and $F(t)=-1$ when $t\in[t_{I}+\tau_{+},t_{I}+\tau_{+}+\tau_{-}]$
with $\tau=\tau_{+}+\tau_{-}$. 
Our protocols starts at $t=0$ and therefore $t_{I}$ defines the initial waveform of the control field. $t_{I}=-\frac{1}{2}\tau_{+}$ ($t_{I}=0$) corresponds to a symmetric (asymmetric) control protocol. To take into account that instataneous switching of the control field could be difficult in
experiments, $F(t)$ can have a transition time $\tau_{s}$ during the switching,  see Fig.~\ref{fig:Fig01}(d). However,  as we will demonstrate in the simulations of this work, as long as  $\tau_{s}$ is not too large, the effect of non-zero $\tau_{s}$ is negligible. For this reason, in the following theoretical analysis, we assume $\tau_{s}=0$. We will see, using uneven durations of the positive and negative drive, i.e., $\tau_{+}\neq \tau_{-}$, can lead to stronger coupling and signal responses.

The condition for the resonance $\delta_{\phi}=0$
gives 
\begin{equation}
\omega_{n}=k_{D}\nu+r_{D}(1-k_{D})\Omega,\label{eq:ResonanceEq}
\end{equation}
where the ratio $r_{D}=(\tau_{+}-\tau_{-})/\tau$ and the frequency
$\nu=2\pi/\tau+r_{D}\Omega$. Note that the resonance frequencies
are different from the frequency, $2\pi/\tau$, of the periodic driving.
Under the resonance condition Eq.~(\ref{eq:ResonanceEq}), there is
coherent coupling of spins, see Fig.~\ref{fig:Fig01}(e). Equation
(\ref{eq:ResonanceEq}) for $k_{D}=0$ corresponds to the resonance
to a spin with $|\omega_{n}|=|r_{D}\Omega|<\Omega$, which is not our
target of control. When $k_{D}=1$ we achieve a spin-spin resonance
at $\omega_{n}=\nu$, which can be much larger than $\Omega$.

Under the resonance condition Eq.~(\ref{eq:ResonanceEq}) the strength $|J|$ has a large value, which is invariant with respect to $t_{I}$ becasue $t_{I}$ only changes the phase factor of $J$. For the case of symmetric control which has $t_{I}=-\frac{1}{2} \tau_{+}$, we obtain  
\begin{equation}
J=4(-1)^{k_{D}}\Omega\sin[\frac{1}{4}(1+r_{D})(\omega_{n}-\Omega)\tau]/[(\omega_{n}^{2}-\Omega^{2})\tau].
\end{equation}
One can maximize the signal for $k_{D}\neq0$ by maximization of $J$.
On this regard we choose $r_{D}=\Omega/\nu$, which implies $\tau_{\pm}=\pi/(\nu\mp\Omega)$
and $g=(2\Omega)/(\pi\nu)$ when $k_{D}=1$.

\begin{figure}[b]
\centering \includegraphics[width=1\columnwidth]{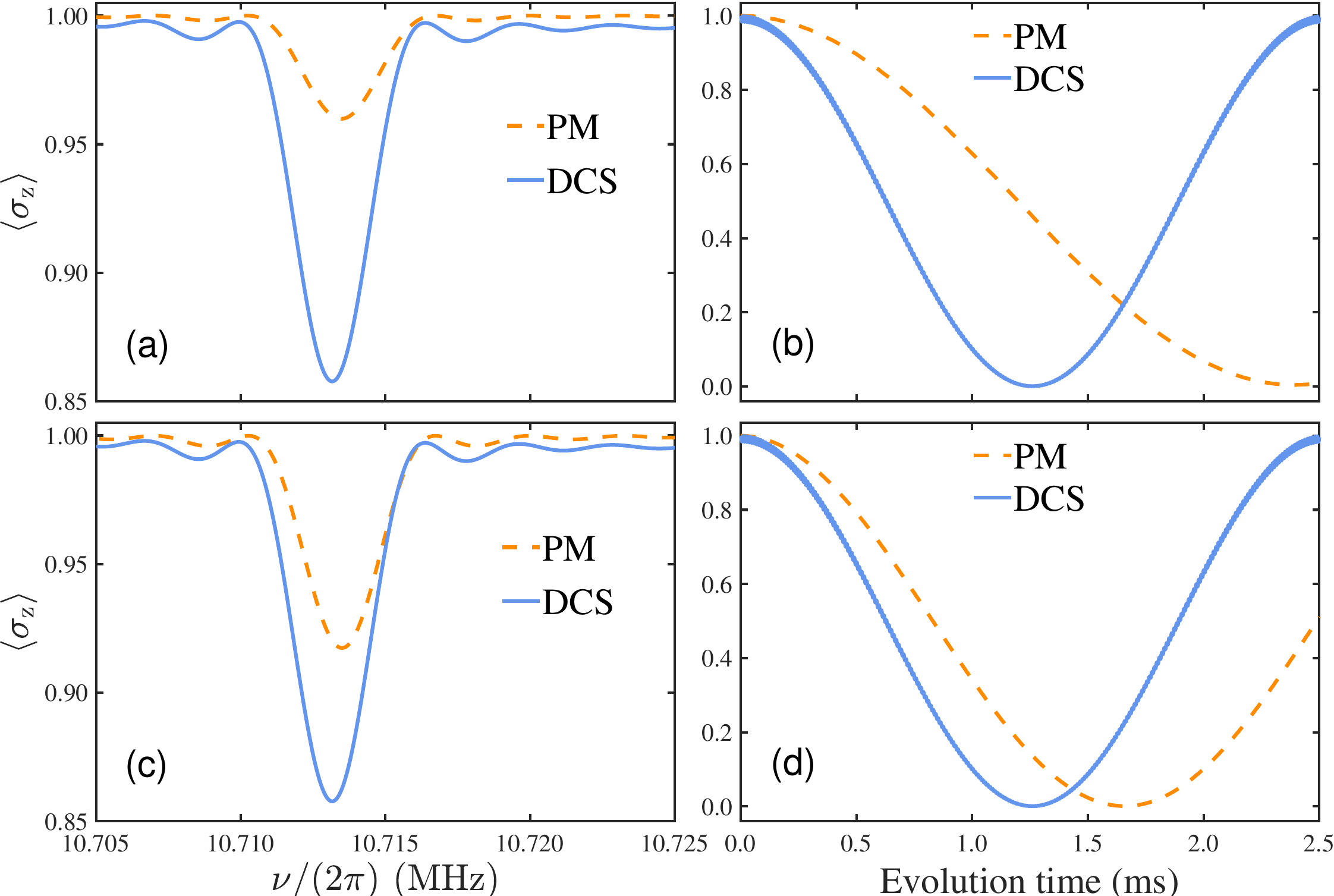} \caption{Polarization signals $\langle\sigma_{z}\rangle$ obtained by using
our DCS (blue solid line)  and those obtained
by the phase modulation (PM) protocol proposed in Refs.~\citep{casanova2019modulated,aharon2019quantum}
(orange dashed line). (a) Spectral response of the signal. The Rabi
frequency of our protocol is $\Omega=2\pi\times1$ MHz. The Rabi frequency
for the PM protocol is $\Omega_{0}+e^{i\theta^{\prime}}\Omega_{1}$,
where $\theta^{\prime}$ is periodically modulated with the values
$0$ and $\pi$ with an interval of $\tau/2$; and we choose the typical
values of $\Omega_{0}=\Omega_{1}=2\pi\times0.5$ MHz as in Ref. \citep{casanova2019modulated}.
Both protocols have the same maximal Rabi frequency. For comparison
the sensing time $T=0.308$ ms is also the same for both protocols.
(b) The signal as a function of the total sensing time $T$ when the
frequency $\nu$ is set to the resonance point in (a). (c) {[}(d){]}
As (a) {[}(b){]}. In (c) and (d) $\Omega_{0}=\Omega_{1}\approx(1/\sqrt{2})\Omega$
for the PM protocol such that its average power \citep{casanova2019modulated}
is the same as our protocol. $t_I=-\frac{1}{2}\tau_{+}$ and $\tau_s=0$ for all plots. The protocol proposed in this work provides
stronger sensing signals \wzy{than previous methods}.}
\label{fig:Fig02}
\end{figure}

\section{Low-power quantum sensing of nuclear spins \wzy{and DNP}}

\begin{figure}[b]
\centering \includegraphics[width=1\columnwidth]{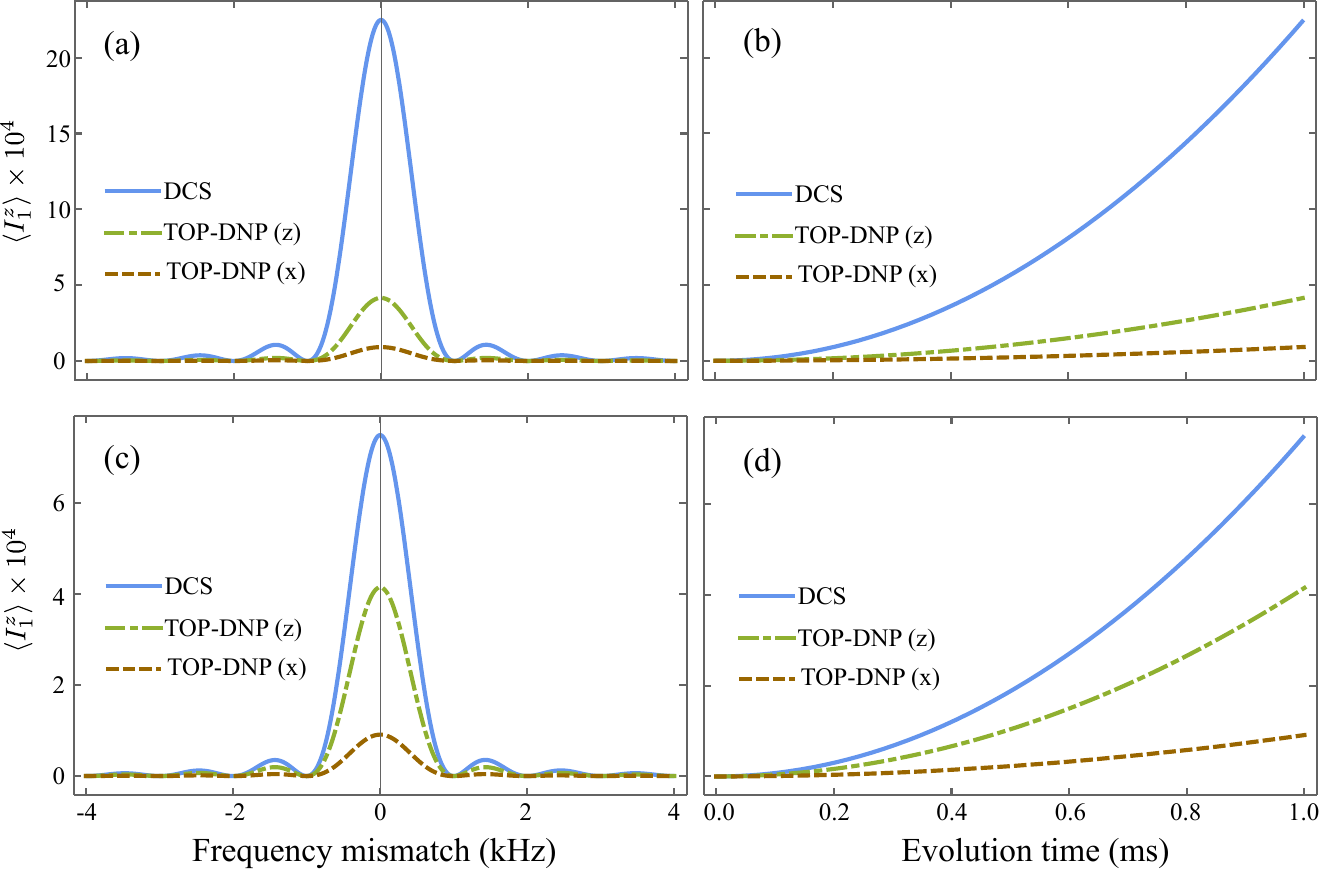} \caption{\wzy{Application to DNP. (a) Blue solid line is the nuclear polarization
after the control of our DCS protocol for a total
time $T\approx 1$ ms, when the nuclear spin is in a high temperature thermal state. The green dash-dotted (brown dashed)
line is the result when one uses the TOP-DNP protocol with the initial electron spin state initialized to be parallel (to be perpendicular) to the direction of the magnetic field of 0.35~T along the $z$ axis, using the optimized parameters such that the microwave pulses have a length $\tau_p=56$ ns and a tunable frequency detuning and are separated by a delay $d=28$ ns (see \citep{tan2019time}). Both protocols have the same maximal value of the Rabi frequency $2\pi\times 2$ MHz. (b) The signal as a function of the total polarization time $T$ when the frequency mismatch vanishes in (a), that is, when $\nu=\omega_{n}$ for the DCS protocol and $\omega_{\text{m}}+\omega_{\text{eff}}=\omega_{n}$, where $\omega_{\text{m}}=2\pi/(\tau_p+d)$ and $\omega_{\text{eff}}$ is the effective field~\citep{tan2019time} for the TOP-DNP protocol.  (c) {[}(d){]}
As (a) {[}(b){]}, but with a reduced Rabi frequency of the DCS protocol, ensuring both DCS and TOP-DNP have the same average power. DCS provides much higher nuclear polarization. Here $\tau_s=0$ and $t_I=-\frac{1}{2}\tau_{+}$ for DCS. }}
\label{fig:FigDNP} 
\end{figure}

We apply our method to an NV center with its surrounding nuclear spins,
which is relevant to quantum information processing, quantum sensing,
and nano-scale NMR, and nuclear hyperpolarization. The Hamiltonian
of an NV center electronic spin and its nearby nuclear spins under
a strong magnetic field $-B_{z}$ along the NV symmetry axis reads
$H^{\prime}=DS_{z}^{2}+\gamma_{e}B_{z}S_{z}+\sum_{j}\gamma_{j}B_{z}I_{z}+S_{z}\sum_{j}(A_{j}^{x}I_{j}^{x}+A_{j}^{z}I_{j}^{z})+H_{c}^{\prime}$,
where $D=(2\pi)\times2.87$ GHz is the NV zero-field splitting, $S_{z}=|1\rangle\langle1|-|-1\rangle\langle-1|+0|0\rangle\langle0|$
for the NV electron spin, $I_{j}^{\alpha}$ ($\alpha=x,y,z$)
are the spin operators for the $j$-th nucleus, and $\gamma_{e}$
and $\gamma_{j}$ are the gyromagnetic ratios for the NV electron
and the nuclear spins, respectively. The components of the hyperfine
coupling $A_{j}^{x}$ and $A_{j}^{z}$ are much smaller than $\gamma_{j}B_{z}$
because of the strong magnetic field $B_{z}$. A microwave control
field with a frequency $\omega_{{\rm mw}}$ applied on the NV center
realizes the control Hamiltonian $H_{c}=\sqrt{2}\Omega F(t)\cos{(\omega_{{\rm mw}}t)}S_{x}$,
where $\Omega F(t)$ is the Rabi frequency of the control field with $F(t)$ being the modulation illustrated in Fig.~\ref{fig:Fig01}(d).

To select two NV electron spin levels $m_{s}=0$ and, say, $m_{s}=1$
to form an NV qubit, we set the microwave frequency to the energy
splitting between the qubit levels. In this manner, we neglect the
state $|-1\rangle$ because there is no transition to it. Moving to
a rotating frame with respect to $H_{0}=DS_{z}^{2}+\gamma_{e}B_{z}S_{z}$,
we get the new Hamiltonian 
\begin{equation}
H=\sum_{j}\gamma_{j}B_{z}I_{z}+|1\rangle\langle1|\sum_{j}(A_{j}^{x}I_{j}^{x}+A_{j}^{z}I_{j}^{z})+H_{c},\label{H}
\end{equation}
where $H_{c}=F(t)\frac{\Omega}{2}(|1\rangle\langle0|+\text{H.c.}$.
Then, with $|+\rangle=\frac{1}{\sqrt{2}}(|1\rangle+|0\rangle$,
$|-\rangle=\frac{1}{\sqrt{2}}(|1\rangle-|0\rangle$, the Pauli operators
$\sigma_{z}=|1\rangle\langle0|+|0\rangle\langle1|=|+\rangle\langle+|-|-\rangle\langle-|$
and $\sigma_{x}=|+\rangle\langle-|+|-\rangle\langle+|$, and the identity
operator $I=|1\rangle\langle1|+|0\rangle\langle0|$ for the NV electron
qubit, Eq.~(\ref{H}) can be written as 
\begin{equation}
H=\omega_{e}(t)\frac{\sigma_{z}}{2}+\sum_{j}\omega_{n,j}I_{j}^{z}+H_{I},\label{15}
\end{equation}
where the qubit energy splitting $\omega_{e}(t)=\Omega F(t)$ is bounded
with the maximal absolute value $\Omega$ because
of experimental limitation, and the nuclear spin frequency $\omega_{n,j}=\gamma_{j}B_{z}+\frac{1}{2}A_{z,j}$
for the $j$th spin is shifted by the hyperfine component $A_{z,j}$,
which is different for different nuclear spins because the electron-nuclear
coupling is position dependent. Here the electron-nuclear interaction
$H_{I}=\frac{1}{2}\sigma_{x}\sum_{j}(A_{j}^{x}I_{j}^{x}+A_{j}^{z}I_{j}^{z})$
in the rotating frame of $\omega_{e}(t)\frac{\sigma_{z}}{2}+\sum_{j}\omega_{n,j}I_{j}^{z}$
becomes 
\begin{align}
\tilde{H}_{I}(t) & =\sum_{j}\frac{1}{4}A_{j}^{z}I_{j}^{z}(\sigma_{+}e^{i\int_{0}^{t}\omega_{e}dt^{\prime}}+{\rm h.c.})\nonumber \\
 & +\sum_{j}\frac{1}{4}A_{j}^{x}(\sigma_{+}I_{j}^{-}e^{i\int_{0}^{t}(\omega_{e}-\omega_{n,j})dt^{\prime}}+{\rm h.c.})\nonumber \\
 & +\sum_{j}\frac{1}{4}A_{j}^{x}(\sigma_{+}I_{j}^{+}e^{i\int_{0}^{t}(\omega_{e}+\omega_{n,j})dt^{\prime}}+{\rm h.c.}),
\end{align}
where $\sigma_{\pm}=\frac{1}{2}(\sigma_{x}\pm i\sigma_{y})$ and $I_{j}^{+}=I_{j}^{x}\pm iI_{j}^{y}$.

\begin{figure}[b]
\centering \includegraphics[width=1\columnwidth]{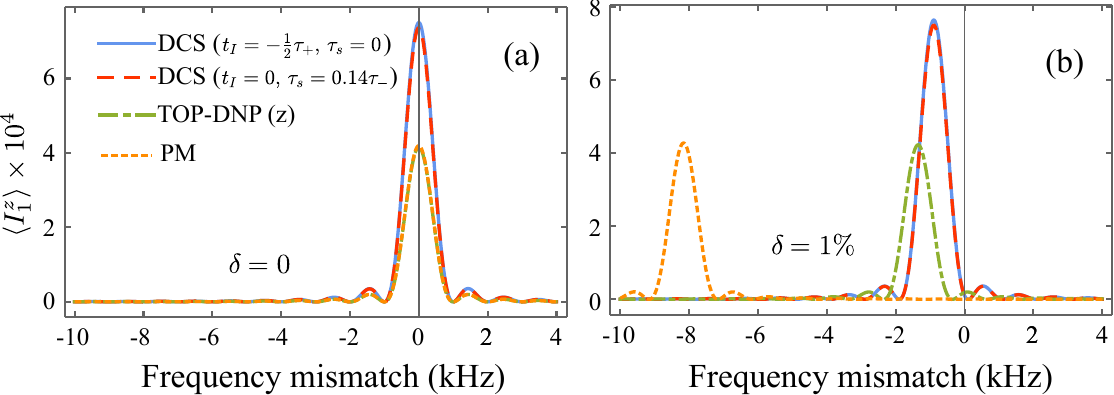} \caption{\wzy{Robustness of the DCS, PM, and TOP-DNP protocols. (a) The nuclear polarization obtained by the protocols without control errors, using the same parameters in  Fig.~\ref{fig:FigDNP}(c). The PM protocol has the same average power as other protocols. The result of DCS with non-instantaneous switching ($\tau_s=0.14 \tau_{-}$) and a different $t_I$ is also shown by a red dashed line. (b) As (a), but by adding a relative error $\delta=1\%$ to the amplitude of the control field. }}
\label{fig:FigRobust} 
\end{figure}

We consider the regime that the nuclear Zeeman energies $\gamma_{j}B_{z}$
is much larger than the Rabi frequency $\Omega$ reachable in experiments,
e.g., due to the low power of the microwave field at the NV center
and the strong magnetic field $B_{z}$. For this situation, the Hartmann-Hahn
condition can not be met for electron-nuclear coupling, see Fig.~\ref{fig:Fig01}(b).
However, according to the resonance condition Eq.~(\ref{eq:ResonanceEq})
for the dynamic phase Eq.~(\ref{eq:phi}), the coupling is preserved
if the  \wzy{frequencies of the} nuclear spin meet the resonance condition. For
example, if only one nuclear spin (say, $j=1$) meets the resonance
condition, we can approximate the interaction by the effective Hamiltonian
in the leading-order (say, by using the Magnus expansion)
\begin{equation}
\tilde{H}_{I}\approx\frac{\Omega}{2\pi\nu}A_{1}^{x}\sigma_{+}I_{1}^{-}+{\rm h.c.},\label{eq:HSingle1}
\end{equation}
when $\omega_{n,1}=\nu$ or 
\begin{equation}
\tilde{H}_{I}\approx\frac{\Omega}{2\pi\nu}A_{1}^{x}\sigma_{+}I_{1}^{+}+{\rm h.c.},\label{eq:HSingle2}
\end{equation}
when $\omega_{n,1}=\nu-2\Omega^{2}/\nu$. Equation (\ref{eq:HSingle1})
or (\ref{eq:HSingle2}) can be used to detect and control single nuclear
spins, e.g., for two-qubit gates between the NV and only one nucleus.

To demonstrate the superior performance of our DCS method in low-power
control, we perform numerical simulations for an NV center and a weakly
coupled $^{13}$C nucleus with $A_{1}^{x}=2\pi\times13.42$ kHz and
$A_{1}^{z}=2\pi\times17.09$ kHz. A strong magnetic field $B_{z}=1$
T gives a strong Zeeman energy $\approx2\pi\times11$ MHz for the
$^{13}$C nuclear spin, which is much larger than the Rabi frequency
of the NV electron spin in our simulations. In the simulations, we
assume that before the protocol the initial state of $^{13}$C nuclear
spin is in a thermal state $\rho_{n}\approx I/2$ and the NV electron
spin is initially prepared in the eigenstate $|+\rangle$ of $\sigma_{z}$.
After a time $T$ of the control, the density matrix of the electron
and nuclear spins becomes $\rho(T)$. As shown in Fig.~\ref{fig:Fig02},
the polarization $\langle\sigma_{z}\rangle={\rm Tr}[\rho(T)\sigma_{z}]$
of the NV electron spin changes significantly at the resonance peak of $\nu=\omega_{n}\approx2\pi\times10.713$
MHz, transferring the electron spin polarization to the nuclear
spin. At $\nu=\omega_{n}$ the signal $\langle\sigma_{z}\rangle=\cos^{2}(\frac{\Omega}{2\pi\nu}A_{1}^{x}T)$.
The realized electron-nuclear coupling is more stronger than the previous
protocol proposed in Refs. \citep{casanova2019modulated,aharon2019quantum},
as shown in Fig.~\ref{fig:Fig02}.

\wzy{We note that our DCS protocol also provides a significant enhancement for DNP. To demonstrate its superior performance, we compare DCS with the time-optimized pulsed dynamic nuclear polarization (TOP-DNP) protocol recently developed in~\citep{tan2019time}, which was shown to perform much better than the traditional nuclear spin orientation via electron spin locking (NOVEL)~\citep{tan2019time}. A TOP-DNP sequence is composed of a train of microwave pulses of a length $\tau_p$ separated by a delay $d$ between the pulses~\citep{tan2019time}. In Fig.~\ref{fig:FigDNP} we consider the polarization of a $^{1}$H spin using an electron spin.  The $^{1}$H spin with $A_{1}^{x}=2\pi\times 0.5$ kHz and
$A_{1}^{z}=2\pi\times 0.5$ kHz under a magnetic field $0.35$~T has a Larmor frequency $\omega_{n}/(2\pi)\approx14.9$ MHz much larger than the Rabi frequency. We use the optimized parameters of TOP-DNP in~\citep{tan2019time}. As shown in Fig.~\ref{fig:FigDNP} our DCS still gives a much higher nuclear spin polarization than the high-performance TOP-DNP protocol. 
}

\wzy{We further compare DCS, PM, and TOP-DNP protocols in Fig.~\ref{fig:FigRobust}, where 
we can see that the DCS protocol gives the same signal response for different values of $t_I$ and when $\tau_{s}$ is not too large. 
The signal strengths of the PM and TOP-DNP protocols are similar and much weaker than that of DCS. 
We compare the protocols when the control field has an error such that the Rabi frequency is changed from
the ideal one $\Omega(t)$ to $(1+\delta)\Omega(t)${]}. A comparison of Fig.~\ref{fig:FigRobust}(a) and (b)
shows that DCS is less sensitive to this error. This robustness is a consequence of the resonance condition
Eq.~(\ref{eq:ResonanceEq}), which implies that an error $\delta\Omega$ of the Rabi frequency only shift
the resonance frequency by a smaller amount $r_D\delta\Omega$ because $r_{D}<1$. The factor $r_{D}\ll1$ for low-power driving $\Omega\ll\omega_n$.  
}

\section{Conclusion}

We have shown that uneven modulation of the dynamic phases via dynamical switching of the energy of a quantum
system, e.g., in a dressed state picture, can induce resonant response
with its surrounding quantum system even through their energy scales
are different. We have applied this idea to achieve low-power quantum
sensing of single nuclear spins \wzy{and DNP} by an electron spin, even through
the available electron spin Rabi frequency of the control field is
much weaker than the frequencies of nuclear spins. Our results show
that our protocol provides much better power efficiency and stronger
sensing signals \wzy{and DNP} than previous power-efficient methods~\wzy{\citep{casanova2019modulated,aharon2019quantum,tan2019time}}.
Our protocol would have useful applications in nanoscale NMR for samples
that are sensitive to heating by control field, in quantum information
processing with high-frequency nuclear spins, as well as other applications
not specific to NV centers.
\begin{acknowledgments}
This work was supported by National Natural Science Foundation of
China (Grant No. 12074131) and the Natural Science Foundation of Guangdong
Province (Grant No. 2021A1515012030).

S.X. and C.X. contributed equally to this work. 
\end{acknowledgments}

%\appendix*
%\section{Appendixes}
%This is an appendix section.

%\bibliography{reference}


\end{document}